\documentclass[10pt,conference]{IEEEtran}
\pdfoutput=1
\IEEEoverridecommandlockouts

\usepackage[english]{babel}
\usepackage[protrusion,expansion]{microtype}

\usepackage{graphicx}
\usepackage{xcolor}
\usepackage{tikz}
\usepackage{booktabs}
\usepackage{listings}
\lstdefinelanguage{diff}{
	sensitive=true,
	morecomment=[f][\color{gray}][0]{diff},
	morecomment=[f][\color{gray}][0]{index},
	morecomment=[f][\color{blue}][0]{@@},
	morecomment=[f][\color{magenta}][0]{***},
	morecomment=[f][\color{violet}][0]{!},
	morecomment=[f][\color{red!60!black}][0]-,
	morecomment=[f][\color{green!40!black}][0]+,
	morecomment=[f][\color{magenta}][0]{---},
	morecomment=[f][\color{magenta}][0]{+++},
	morecomment=[f][\color{gray}][0]{Binary},
	morecomment=[f][\color{gray}][0]{Only},
	morecomment=[f][\color{gray}][0]{old},
	morecomment=[f][\color{gray}][0]{new},
	morecomment=[f][\color{gray}][0]{rename},
	morecomment=[f][\color{gray}][0]{similarity},
	morecomment=[f][\color{gray}][0]{deleted},
	morecomment=[f][\color{magenta}][0]{***************},
	morecomment=[f][\color{red!60!black}][0]<,
	morecomment=[f][\color{green!60!black}][0]>,
	morecomment=[f][\color{blue}][0]{0},
	morecomment=[f][\color{blue}][0]{1},
	morecomment=[f][\color{blue}][0]{2},
	morecomment=[f][\color{blue}][0]{3},
	morecomment=[f][\color{blue}][0]{4},
	morecomment=[f][\color{blue}][0]{5},
	morecomment=[f][\color{blue}][0]{6},
	morecomment=[f][\color{blue}][0]{7},
	morecomment=[f][\color{blue}][0]{8},
	morecomment=[f][\color{blue}][0]{9},
}[comments]

\usepackage{subcaption}
\usepackage{float}
\usepackage[T1]{fontenc}

\usepackage{amsmath,amssymb,amsfonts,amsthm}
\usepackage{MnSymbol}
\usepackage[maxbibnames=9, minbibnames=3, backend=biber, doi=false, url=false, isbn=false]{biblatex}
\bibliography{literature}

\title{The List is the Process: Reliable Pre-Integration Tracking
  of Commits on Mailing Lists}
\author{
\IEEEauthorblockN{
       Ralf Ramsauer\IEEEauthorrefmark{1},
       Daniel Lohmann\IEEEauthorrefmark{2} and
       Wolfgang Mauerer\IEEEauthorrefmark{1}\IEEEauthorrefmark{3}}
\IEEEauthorblockA{
       \IEEEauthorrefmark{1}Technical University of Applied Sciences Regensburg\\
       \IEEEauthorrefmark{2}University of Hanover\\
       \IEEEauthorrefmark{3}Siemens AG, Corporate Technology, Munich\\
       ralf.ramsauer@othr.de,
       lohmann@sra.uni-hannover.de,
       wolfgang.mauerer@othr.de
\thanks{This work was supported by Siemens AG, Corporate Research, the iDev40
project and the German Research Council (DFG) under grant no. LO 1719/3-1. The
iDev40 project has received funding from the ECSEL Joint Undertaking (JU) under
grant no.~783163.  The JU receives support from the European Union’s
Horizon~2020 research and innovation programme. It is co-funded by the
consortium members, grants from Austria, Germany, Belgium, Italy,
Spain and Romania.}
}
}

\begin{document}
\maketitle
\begin{abstract}
	A considerable corpus of research on software evolution focuses on mining
changes in software repositories, but omits their pre-integration history.

We present a novel method for tracking this otherwise invisible evolution of
software changes on mailing lists by connecting all early revisions of changes
to their final version in repositories. Since artefact modifications on mailing
lists are communicated by updates to fragments (i.e., patches) only,
identifying semantically similar changes is a non-trivial task that our
approach solves in a language-independent way. We evaluate our method on
high-profile open source software (OSS) projects like the Linux kernel, and
validate its high accuracy using an elaborately created ground truth.

Our approach can be used to quantify properties of OSS development processes,
which is an essential requirement for using OSS in reliable or safety-critical
industrial products, where certifiability and conformance to processes are
crucial. The high accuracy of our technique allows, to the best of our
knowledge, for the first time to quantitatively determine if an open
development process effectively aligns with given formal process requirements.

\end{abstract}
\begin{figure*}
	\includegraphics[width=1\textwidth]{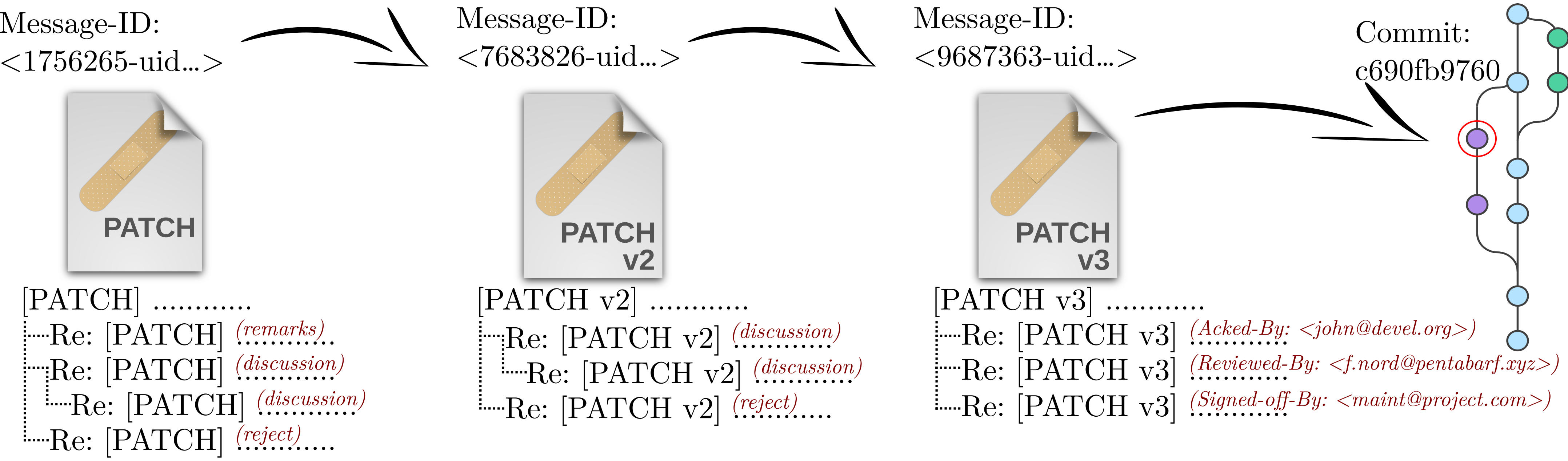}
	\caption{Typical workflow: A patch gets resubmitted and improved for two times, before its integration}
	\label{fig:workflow}
\end{figure*}

\section{Introduction}
Software patches may have come a long way before their final
integration into the official branch (known as \emph{mainline} or
\emph{trunk}) of a project. There are many possible ways of
integration. Among others, the origin of a patch can be a merge from
other developers' repositories (i.e., integration of branches or
patches from foreign repositories), pull requests on web-based
repository managers such as Github or Gitlab, vendor specific
patch stacks, or mailing lists (MLs).

Especially MLs have been in use for software development processes for
decades~\cite{erenkrantz2003release}. They have a well-known interface (plain
text emails), and come with an absolute minimum of tool requirements (i.e., a
mail user agent). Because of their simplicity, scalability, reliability and
interface robustness, they are still widely used in many open source software
(OSS) projects. In particular, mailing lists are a core infrastructure
component of long-lasting OSS projects such as low-level systems software
(e.g., QEMU, U-Boot, GRUB, etc.), operating systems (e.g., the Linux kernel) or
foundations (e.g., Apache, GNU): Mailing lists form the backbone of their
development processes~\cite{hertel2003motivation}. They are not only used to
ask questions, file bug reports or discuss general topics, but implement a
patch submit-review-improve strategy for stepwise
refinement~\cite{wirth-stepwise} that is typically iterated multiple times
before a patch is finally integrated to the repository
(cf.~Figure~\ref{fig:workflow}).

Therefore, MLs contain a huge amount of information on the pre-integration
history of patches. A commit in a repository may be the outcome of that
process, while all intermediate steps leave no direct traces in the repository.
Mailing lists allow us to analyse development history and code evolution, but
also enable us to inspect reviewing and maintenance processes. They further
allow inferring organisational~\cite{core-peripheral} and
socio-technical~\cite{bird2006mining, herbsleb2007global, valetto2007using}
aspects of software development. This all is possible because MLs contain
information on interactions between developers.

Nowadays, open source components are routinely deployed in industrial fields,
and their use is increasingly explored in safety-critical or mixed-criticality
appliances~\cite{burns2013mixed}, such as medical devices or in automotive
products. Especially for core components of a system that implement
business-wise non-differentiating features such as the system-software stack or
middleware, OSS provides adequate solutions that have already proved to be
reliable in other non-critical application domains.

However, non-functional aspects like evidences of quality assurance are also a crucial factor for
industry. Deployment of software in safety-critical environments requires
conformance with international standards, such as ISO~26262~\cite{iso26262},
IEC~61508~\cite{iec61508} or IEC~62304~\cite{iec62304}. This demands certified
development processes that implement high standards regarding traceability and
auditability of all development decisions, including code writing, reviewing,
deployment, and maintenance activities (the rationale for strict process
compliance is to achieve and prove high product quality).

Compared to conventional, orthodox proprietary industrial software, OSS
exhibits different dynamics~\cite{mauerer2013open}, and often requires
fundamentally different development processes~\cite{corbet2011process} because
of project size and a high number of massively geo-dispersed stakeholders.
Because of this nature of OSS, projects do not necessarily meet certification
criteria~\cite{bird2008latent}.

Nevertheless, vendors across different industrial sectors share similar
concerns on the use of OSS components~\cite{agl, cip}: OSS projects are
community driven. Hence, their established processes can only be applied to a
certain degree. Quantitative ex-post analyses of processes are required to
investigate conformance. Statistical methods are necessary to judge the
applicability of OSS components in different scenarios. This makes it possible
to reconstruct process operations, and use them to draw conclusions on
processes with quantitative software-engineering techniques. However, how to do
this is an unsolved issue in industry~\cite{osadl-sil, leitner2017open}.

To assess non-formal OSS development processes, mapping patches on mailing
lists to repositories is a key requirement, because the mails contain the
facts: They are the artefacts of the development process. Together with the
outcome of the process---the repository---, this forms a solid base for further
analysis. Patches that appear on mailing lists are manually selected
(\emph{cherry-picked}) by the maintainer before integration into the
repository. They are also routinely combined (\emph{squashed}) and modified
(\emph{amended}) on-the-fly, which is convenient for developers, but
complicates tracking. Either way, a direct connection between the history on
the mailing list and the repository commit is lost in the
process~\cite{bird2007detecting}.

We present a method accompanied by comprehensive automated tool
support\footnote{Published under the GPLv2 license at~\url{https://github.com/lfd/PaStA}}
that allows us (a) to track several revisions of a patch on a mailing list, and
(b) to map those patches on the list to upstream commit hashes, if the patch
was integrated. We identify and formalise the problem as cluster analysis, and
provide an in-depth evaluation of our and other approaches. Both problems are
reduced to finding similar patches. We quantify the accuracy of the approaches
with elaborate external validation measurements based on a ground truth in
Section~\ref{sec:evaluation}. We claim the following contributions:

\begin{itemize}
	\item A novel, highly accurate methodology to reconstruct the missing
		link between mailing lists and repositories on noisy real-world
		data.
	\item A precise formalisation of the problem, together with a
		previously unavailable elaborate external validation of our
		algorithm based on a proper ground truth, together with a
		qualitative evaluation of other approaches.
	\item An industry-grade, fully published and extensible framework that
		allows for further in-depth analyses and scales to handle the world's
		largest software development projects.
\end{itemize}

Results of the evaluation of the Linux kernel and its principal ML underline
the high accuracy of our approach.

\section{Related Work I}
A patch consists of an informal commit message that describes the changes of
the patch in natural language, and annotations of the modifications to files of
a project. First and foremost, patches modify source code, but also
documentation, build system, tools and any other artefacts of a project. A
single patch may modify several files. Within the context of a file,
\emph{chunks} (also known as hunks) are segments that describe changes to a
certain area within a file. Figure~\ref{fig:example} illustrates the typical
structure of patches on the ML (a, b) and in the repository (c). We need to
find similar patches to track patch evolution.

Jiang, Adams and German~\cite{jiang2013will} present a coarse-grained
checksum-based technique for mapping emails that contain patches to commits.
After trimming whitespaces they calculate MD5 hashes over chunks of the patch.
Two patches are considered similar if they have at least one checksum in common
(i.e., share one equivalent chunk).

In another work~\cite{german-tracing}, the authors refine their technique and
present further approaches: A plus-minus-line-based technique and a
clone-detection-based technique. The plus-minus-line--based technique weights
the fraction of equivalent lines of two patches. This includes insertions (+)
and deletions (-). The clone-detection--based technique incorporates
CCFinderX~\cite{bettenburg2012using}, a code-clone detector. They evaluate
their three techniques, and conclude that the plus-minus-line--based technique
is performing best. This evaluation is based on the F-Score that depends on the
precision and recall of the actual algorithm. In contrast to measuring the
precision, the F-Score requires a ground truth for determining the recall. As a
ground truth is hard to obtain, authors use the concept of \emph{relative
recall} that provides a qualitative approximation.

We presented a method and a tool to identify similar patches in different
branches of a repository~\cite{ramsauer2016observing}.  They use their method
to quantify integration efforts of huge software forks, like the PREEMPT\_RT
real-time patch for the Linux kernel, or hardware-vendor--specific forks of the
Linux kernel. The problem is to find patches that first appeared in a
development branch, and were later applied to the master branch of the project.
Yet, this work misses a proper quantitative evaluation, and only operates on
commits within a repository.

\section{Research Methods}

From an analytical standpoint, the downside of patch submission on mailing lists
is asynchronicity, as there is no direct connection between the mailing list
and the software repository. Maintainers manually integrate patches from the
list and commit them to the repository. This process is typically assisted by
tools provided by the version control system.\footnote{e.g., git am (apply mail
from mailbox) or git cherry-pick (apply the changes introduced by some existing
commits)} During this process, the connection of the email with the patch
(identified by the unique Message-ID header of the mail) and the commit in the
repository (usually identified by a commit hash) is lost.

Other difficulties are contextual divergences and textual
differences~\cite{bird2007detecting}. The commit in the repository may
significantly vary from the patch on the mailing list, as other patches between
submission and integration might have affected the patch. Additionally,
maintainers may introduce additional changes to the patch.

There is also no connection between several revisions of a patch within the
mailing list. A patch undergoes a certain evolutionary process between
revisions, hence patches of different revisions may significantly differ as
well, while they still introduce the same logical change.

\subsection{Code Submission Workflow}

Independent of the type of submission, a patch $p$ is formally defined as a
2-tuple that consists of a commit message and a diff. While the commit message
informally describes the changes, the diff annotates the actual modifications
(insertions and deletions) surrounded by a few lines of context. Context lines
ease the understandability of the patch for human review. Patches can also
include meta information, such as the author of a patch or the timestamp of its
creation (Author Date). Not all types of patches contain the same set of
metadata. Emails with patches contain several mail headers, while those headers
are removed when the patch is applied to the repository. Repositories, in
contrast, contain information on the exact spatial location of the patch.

Metadata may also change over time~\cite{bird09, german2016mining}; even the
author of a patch may change. Therefore, we intentionally do not consider
metadata in our similarity analysis.

Mapping patches on mailing lists to commits in repositories requires to
understand common workflows in projects~\cite{erenkrantz2003release}: When the
author of a patch wants his or her patches to be integrated in the project,
they need to send their patch or \emph{patch series} to the mailing list of the
project.
\begin{figure}[H]
	\begin{subfigure}{\linewidth}
		\resizebox{1.0\linewidth}{!}{
\begin{lstlisting}[language=diff]
Message-ID: <1338734589-11512-3-git-send-email-tias@ulyssis.org>
Date: Sun,  3 Jun 2012 16:43:04 +0200
To: Discussion and development of BusyBox <busybox.busybox.net>
From: Tias Guns <tias@ulyssis.org>
Subject: [PATCH 2/6] android: use BB_ADDITIONAL_PATH

Signed-off-by: Tias Guns <tias@ulyssis.org>
---
 include/platform.h |    4 ++++
 1 file changed, 4 insertions(+)

diff --git a/include/platform.h b/include/platform.h
index d79cc97..f250624 100644
--- a/include/platform.h
+++ b/include/platform.h
@@ -334,6 +334,10 @@ typedef unsigned smalluint;
 # define MAXSYMLINKS SYMLOOP_MAX
 #endif
 
+#if defined(ANDROID) || defined(__ANDROID__)
+# define BB_ADDITIONAL_PATH ":/system/sbin:/system/bin:/system/xbin"
+#endif
+
 
 /* ---- Who misses what? ------------------------------------ */
 
-- 
1.7.10
\end{lstlisting}
		}
		\caption{\texttt{[PATCH 2/6]} in a series: the author adds some conditional preprocessor definitions}
	\end{subfigure}
	\vskip 10pt
	\begin{subfigure}{\linewidth}
		\resizebox{1.0\linewidth}{!}{
\begin{lstlisting}[language=diff]
Message-ID: <1338734589-11512-4-git-send-email-tias@ulyssis.org>
Date: Sun,  3 Jun 2012 16:43:05 +0200
To: Discussion and development of BusyBox <busybox.busybox.net>
From: Tias Guns <tias@ulyssis.org>
Subject: [PATCH 3/6] android: fix 'ionice', add ioprio defines

patch inspired by 'BusyBox Patch V1.0 (Vitaly Greck)'
https://code.google.com/p/busybox-android/downloads/detail?name=patch_busybox

Signed-off-by: Tias Guns <tias@ulyssis.org>
---
 include/platform.h |    2 ++
 1 file changed, 2 insertions(+)

diff --git a/include/platform.h b/include/platform.h
index f250624..ba534b2 100644
--- a/include/platform.h
+++ b/include/platform.h
@@ -336,6 +336,8 @@ typedef unsigned smalluint;
 
 #if defined(ANDROID) || defined(__ANDROID__)
 # define BB_ADDITIONAL_PATH ":/system/sbin:/system/bin:/system/xbin"
+# define SYS_ioprio_set __NR_ioprio_set
+# define SYS_ioprio_get __NR_ioprio_get
 #endif
 
 
-- 
1.7.10

\end{lstlisting}
		}
		\caption{\texttt{[PATCH 3/6]} in a series: the author adds further definitions under the same condition}
	\end{subfigure}
	\vskip 10pt
	\begin{subfigure}{\linewidth}
		\resizebox{1.0\linewidth}{!}{
\begin{lstlisting}[language=diff]
commit 3645195377b73bc4265868c26c123e443aaa71c6
Author: Tias Guns <tias@ulyssis.org>
Date:   Sun Jun 10 14:26:32 2012 +0200

    platform.h: Android tweaks: ioprio defines, BB_ADDITIONAL_PATH
    
    Signed-off-by: Tias Guns <tias@ulyssis.org>
    Signed-off-by: Denys Vlasenko <vda.linux@googlemail.com>

diff --git a/include/platform.h b/include/platform.h
index d79cc97..ba534b2 100644
--- a/include/platform.h
+++ b/include/platform.h
@@ -334,6 +334,12 @@ typedef unsigned smalluint;
 # define MAXSYMLINKS SYMLOOP_MAX
 #endif
 
+#if defined(ANDROID) || defined(__ANDROID__)
+# define BB_ADDITIONAL_PATH ":/system/sbin:/system/bin:/system/xbin"
+# define SYS_ioprio_set __NR_ioprio_set
+# define SYS_ioprio_get __NR_ioprio_get
+#endif
+
 
 /* ---- Who misses what? ------------------------------------ */
 
\end{lstlisting}
		}
		\caption{Maintainer squashed both mails to one commit and amended the commit message}
	\end{subfigure}
	\caption{Example of two mails and one commit that were automatically found and linked by our tool}
	\label{fig:example}
\end{figure}

A patch series is a cohesive set of mails that contain several
logically connected patches that, in the big picture, introduce one logical
change that is split up in fine granular steps. Figure~\ref{fig:example} (a)
and (b) show two successive mails in a patch series. The submission of a patch
or patch series is typically tool-assisted by the version control
system.\unskip\kern-2pt\footnote{e.g., git format-patch in combination with git
send-email}

After patches are submitted, reviewers or any subscriber of the list may
comment on them. This is done by starting a free-form textual discussion by
replying to a mail. Inline comments refer to the related code lines.

Concerning change integration, the reviewing process may end up in the
following scenarios: (1) The maintainer decides to integrate (commit) the
patch(es), (2) the maintainer decides to reject the patch(es), (3) the
patch(es) need further improvement and need to be resubmitted to the list. It
is not unusual that (3) is repeated several times. In this case, further
revisions of the patch are typically tagged in the email subject header with
\texttt{[PATCH v<N>]} prefix, where \texttt{<N>} denotes the the revision
round.  This iterative process of resubmitting further revisions of changes
is a fundamental aspect of the development process and makes it necessary that
a patch on a mailing lists must not only be linked to the repository, but also
against other revisions of the patch in order to track its evolution.
Figure~\ref{fig:workflow} illustrates a typical workflow: a patch was resent
two times (v2 and v3), before being integrated to the repository.

Once maintainers decide to accept a patch, they may still amend the commit
message or the code. Depending on the submission process of the project,
maintainers or other persons working on the patch add additional \emph{tags} to
the commit message, such as \texttt{Acked-by: <mail>}, \texttt{Tested-by:
<mail>}, \texttt{Signed-off-by: <mail>} among others.

Reviewers that vote for inclusion of the patch reply to it with a mail that
adds an \texttt{Acked-by}, where \texttt{<mail>} contains the email address of
the person who acknowledged the patch. Anyone who successfully tested a patch
may send their \texttt{Tested-by}. The \texttt{Signed-off-by} tag indicates
that the patch conforms with the Developer's Certificate of Origin\footnote{see
Linux's Documentation/process/submitting-patches.rst}. Maintainers pick up
mails with such tags (i.e., mails \texttt{In-Reply-To} the initial patch) and
append them to the commit message before integration.

A patch on a list may significantly differ from its final version in the
repository, which makes it hard to link them. Figure~\ref{fig:example}
demonstrates the complexity of finding similar patches. This examples contains
two patches that appeared on the mailing list of BusyBox~\cite{bbox}
and the eventual commit in the repository. In this case, the maintainer (Denys
Vlasenko) heavily changed the original patches (authored by Tias Guns) that
were sent to the project's mailing list: He picked up both mails, consolidated
them to one commit (known as \emph{squashing patches}) and additionally changed
the commit message.  During this process, metadata changed as well: the author
date of the commit message is neither related to \texttt{[PATCH 2/6]} nor to
\texttt{[PATCH 3/6]}. Still, both emails are related to the commit in the
repository, and mails and commit were automatically linked by our tool.

The complexity of finding similar patches is aggravated by the fact that
patches are relative to a specific state of the code base, determined by the
commit where the patches base on. When the latter changes between the time a
patch was submitted and it was integrated, as other patches had been applied
 meanwhile, the version control system tools try to (semi-)automatically
adopt the changes, which leads to different context information despite
identical changes. If automatic methods fail, merge conflicts must manually be
solved by humans.

Multiple maintainers may commit the same patch to their own branch. In this
case, a patch occurs multiple times on the master branch of the repository,
once those branches are merged.

Those and other facts~\cite{bird09, german-tracing} underline that similar
patches can not be simply linked against each other by examining their textual
equality.

\subsection{Linking similar patches}
We use and extend the method that we presented in~\cite{ramsauer2016observing}
to work on mailing lists.

Let $\mathcal{C}$ be the set of all patches (commits) in a software repository,
and $\mathcal{M}$ be the set of all patches on a mailing list (mails containing
patches). The universe $\mathcal{U}=\mathcal{M}\cup\mathcal{C}$ forms the set
of all patches.

In its most general form, the informal equivalence relation $S:$
\emph{patches are semantically similar} can be defined as
$S\subseteq\mathcal{U}\times\mathcal{U}$. This covers all
eventualities, including situations like \emph{patch
  committed twice in the repository} or \emph{patch went through
  several rounds of review before integration}.

The algorithm in~\cite{ramsauer2016observing} is able to quantify the
similarity of two patches within a repository by four parameters (explained in
Section~\ref{sec:params}) that influence the sensitivity of the algorithm. It
measures the similarity of two patches
\begin{equation}
	\text{sim}_\text{tf,th,dlr,w}\text{: }\mathcal{U}\times\mathcal{U}\to[0,1]
\end{equation}
where 0 denotes complete dissimilarity (i.e., no commonalities) and 1
denotes complete equivalence on a textual level. Note that symmetry
\begin{equation}
	\forall a,b \in \mathcal{U}:
		\text{sim}_\text{tf,th,dlr,w}(a, b) =
		\text{sim}_\text{tf,th,dlr,w}(b, a)
\end{equation}
and reflexivity
\begin{equation}
	\forall a \in \mathcal{U}:
		\text{sim}_\text{tf,th,dlr,w}(a, a) = 1
\end{equation}
hold.

\begin{figure}[t]
	\hspace*{-40pt}
	\includegraphics{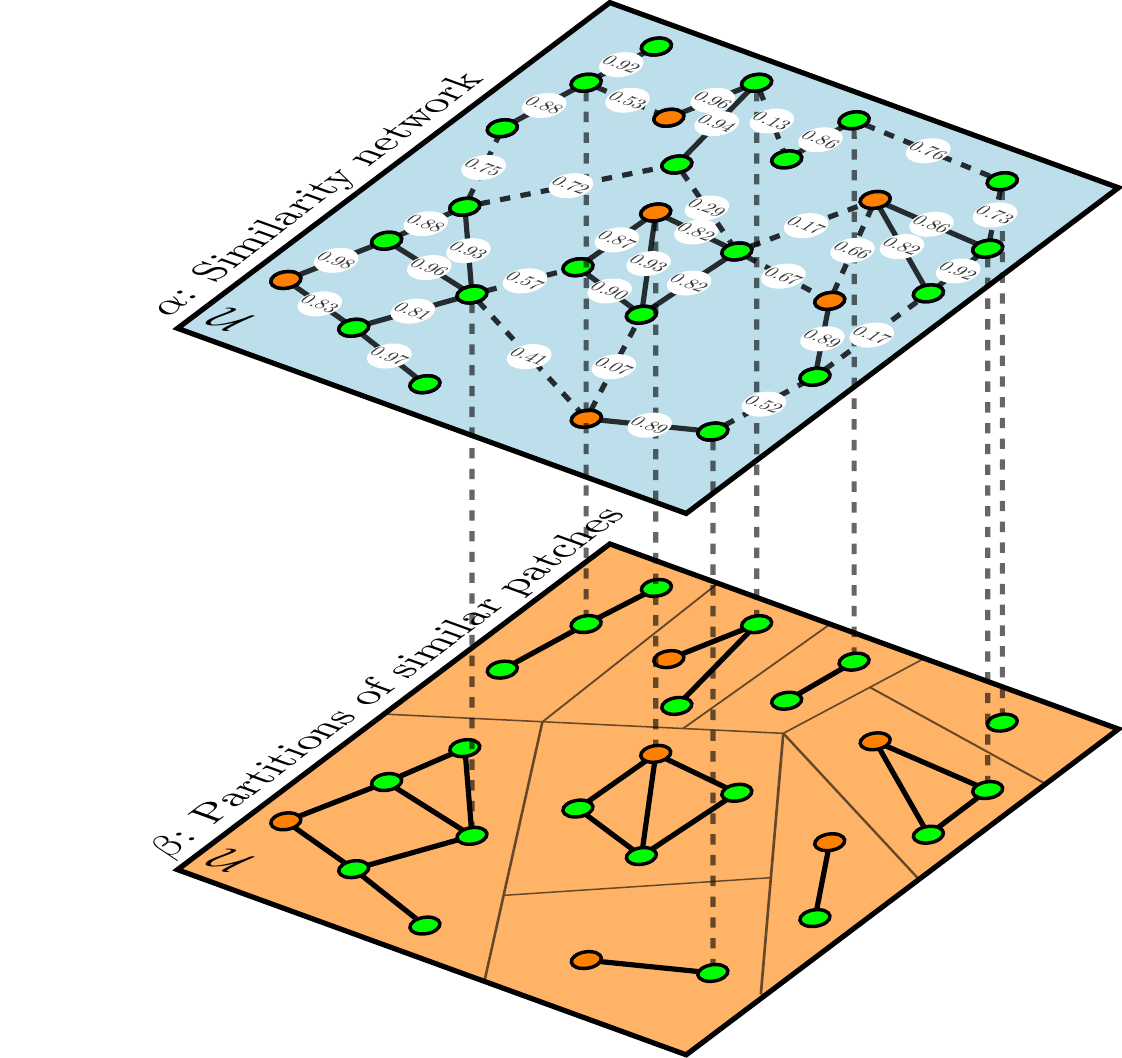}
	\caption{$\alpha$: sim determines the similarity (edge weights) of patches.
		Dashed edges remain below the threshold $\text{ta} = 0.80$. $\beta$:
		Connected components above the threshold form equivalence classes of
		similar patches. Green and orange vertices exemplarily denote patches
		on ML and commits respectively.}
	\label{fig:graph}
\end{figure}

Let $V=\mathcal{U}$ be the set of all vertices of the undirected graph
$G=(V,E)$. Every edge in $E$ connects two patches that exceed the
threshold ta:
\begin{equation}
	E = \{ \{a, b\} \subseteq U | \text{sim}_\text{tf,th,dlr,w}(a, b) > \text{ta} \}
\end{equation}
The connected components of $G$ form subgraphs of similar patches
that divide $\mathcal{U}$ into disjoint partitions. Those partitions induce 
equivalence classes
\begin{equation}
	[x]_S = \{ y \in V | x \leadsto_G y \}
\end{equation}
where $\leadsto_G$ denotes reachability. We use $\sim_S$ to denote the
corresponding equivalence relation, and can use sim to determine all
equivalence classes by pairwise patch comparison in a process that iteratively
merges equivalence classes where the similarity of two patches exceeds a
certain threshold ta (cf.~Figure~\ref{fig:graph}). Section~\ref{sec:reduction}
describes how we overcome resulting combinatorial explosion.

From another perspective, the partition of the equivalence relation $S$ can
also be seen as an unsupervised threshold-based flat clustering of
$\mathcal{U}$~\cite{inforetr}. In Section~\ref{sec:evaluation}, we will use
this fact to evaluate the accuracy of the approach with external evaluation
methods for clusterings.

With this, we reduced the problem of finding clusters of similar patches to a
function sim, which rates the similarity of two patches.
In the following, we will introduce sim, the function that scores the
similarity of two patches, and its set of parameters that control the
sensitivity of the function.

\subsubsection{Rating similarity of two patches}
As mentioned above, patches evolve over time. While the commit message and the
code may change, they still introduce the same logical change. As the commit
message and diff may evolve independently, we calculate two independent scores
that quantify the similarity of the two commit messages and the similarity of
the two diffs ($r_\text{msg}, r_\text{diff} \in [0,1]$). Again, $0$ means no
commonalities while $1$ means equivalence on a textual level.

\paragraph{Similarity of commit messages}
Maintainers may amend or reword commit messages before they integrate the
patch. They can also rearrange or reformat the patch to make it easier to
understand, or to avoid ambiguities. Nevertheless, keywords that are used in
those messages tend to remain the same. Before comparing commit messages, we
remove all tags that were added by maintainers, as they do not appear in the
initial patch. The next step is to tokenise and sort all words in a commit
message. The tokens are separated by whitespaces. We then pairwise compare them
against each other by using the Levenshtein string distance~\cite{levenshtein}.
We select the closest match for each token. The arithmetic mean over all
matches forms the score $r_\text{msg}$. We chose the Levenshtein string
distance together with tokenisation, as it respects restructured messages as
well as minor changes in wording, such as typo fixes.

\paragraph{Similarity of diffs}
Even if code changes or evolves over time, we observed that different versions
of a patch very likely still affect the same code paths and files and use
similar keywords or variable names. We compare diffs in an iterative process. A
single patch may modify several files. When comparing the diff component of two
patches, we only consider changes to files with similar filenames. The
threshold of the Levenshtein similarity for filenames is determined by the
parameter tf, which must be exceeded if the diff of two files is considered for
actual comparison. A diff of a given file may consist of several \emph{hunks},
which describe changes to a certain section within the file. Hunks are
annotated with the line number within the file and a \emph{hunk header} that
describes the context of the change (cf.~Figure~\ref{fig:example}). They
display "the nearest unchanged line that precedes each hunk"~\cite{diff}. We
pairwise compare all hunks of the two diffs against each other, but only
consider hunks with hunk headers that exceed a certain similarity th. Hunks for
which a mapping can not be established are ignored, as the hunk might have been
added or removed in one of the patches. To compare those hunks, we disregard
context lines as they might have changed in the meanwhile, compare insertions
only against insertions, and deletions only against deletions. Therefore, we
again tokenise deletions resp. insertions and use the Levenshtein string
distance to compute a score for the hunk. The arithmetic mean of scores of all
hunks provides the similarity score for the diff, $r_\text{diff}$.

\subsection{Parameters}
\label{sec:params}

The extensive use of string metrics for measuring the similarity of different
parts of a patch opens a wide spectrum for different thresholds of similarity.
Additional parameters (tf, th, dlr, w, ta) investigate the structure of the
patch and control the sensitivity of the comparison.

\paragraph{tf: filename threshold}
A file might have been renamed in the time window between the submission and
acceptance of a patch. As mentioned above, we only consider the pairwise
comparison of files with a similar filename. The filename threshold ($\text{tf}
\in [0,1]$) denotes a similarity threshold for filenames that must be exceeded
if two files shall be considered for comparison.

\paragraph{th: hunk header threshold}
Within a file, the location of a hunk might have moved in the time window
between submission and acceptance of a patch. Either the author moved the
location of the hunk, the upstream location changed or a maintainer moved the
code. Hunk headings try to ease the readability of the patch. Regular
expressions backward-search for anchor lines that will appear in the hunk
heading, such as, e.g., function names. The hunk heading threshold ($\text{th}
\in [0,1]$) denotes the similarity of two hunk headings of hunks that must be
exceeded if two hunks shall be considered for comparison.

\paragraph{dlr: diff-Length ratio}
Similar patches only slightly differ in size. It is unlikely that a patch that
modifies one single line is related to a patch that affects hundreds of lines.
Because of this, patches are considered dissimilar if the diff-length ratio
($\text{dlr} \in [0,1]$), which is the fraction of the number of changed lines
of the smaller patch by the number of lines patched by the bigger patch, is not
exceeded.

\paragraph{w: commit-diff weight}
Since we calculate two independent scores for the commit message and for the
diff, a heuristic factor $w \in [0,1]$ weights the relative importance of
$r_\text{diff}$ to $r_\text{msg}$ and denotes the overall similarity:
\begin{equation}
	\text{sim}_\text{tf,th,dlr,w}(a, b) =
		\begin{cases}
			0\ \text{if}\ \min(a,b)/\max(a,b) < \text{dlr} \\
			w \cdot r_\text{msg}(a, b) + (1-w) \cdot r_\text{diff}(a, b)\ \text{else}

		\end{cases}
\end{equation}

\paragraph{ta: auto accept threshold}
The auto accept threshold ta denotes the required score for patches to be
considered similar. Patches are only considered similar, if
\begin{equation}
	\text{sim}_\text{tf,th,dlr,w}(a, b)\geq \text{ta}
\end{equation}

Section~\ref{sec:evaluation} investigates the significance of the chosen set of
parameters.

The selection of these metrics is based on domain specific expert knowledge of
the Authors, which is provided by participation and contributions in a range of
OSS projects, and during the development of our tool. We observed some
peculiarities of patches that can be used to parameterise the comparison:
\begin{enumerate}
	\item Files may be moved in the repository between submission and
	      acceptance of a patch.
	\item Files in the repository may undergo other changes between
	      submission and acceptance of a patch. This might lead to merge
	      conflicts that have been resolved. Merge conflicts change the
	      context of a patch.
	\item It is unlikely that small patches (e.g., \emph{one-liners}) are
	      related to a huge patch (e.g., feature-introducing patches that
	      add thousands of lines).
	\item Different projects have different maintenance strategies. In some
	      projects, maintainers heavily modify commit messages (see
	      Figure~\ref{fig:example}), in other projects maintainers might
              leave the commit message as it is, but modify the code.
\end{enumerate}

\subsection{Reduction of problem space and clustering patches}
\label{sec:reduction}
The major practical challenge of our approach is scalability. Consider a huge
project like the Linux kernel. Our mailing list archive reaches from
2002-01\,--\,2018-07 and contains $\approx 2.8\cdot 10^6$ mails where
$|\mathcal{M}|\approx 8.5\cdot 10^5$ mails contain patches. The corresponding
upstream range (v2.6.12--v4.18) contains $|\mathcal{C}|\approx 7.6\cdot 10^5$
commits.  This leads to a patch universe of $|\mathcal{U}|\approx 1.6\cdot
10^6$ entries, with a total number of $\binom{|\mathcal{U}|}{2} \approx
1.3\cdot 10^{12}$ pairwise comparisons.

In a preevaluation phase, we drastically reduce the impractical number of
pairwise comparisons. First and foremost, we only consider pairs of patches for
comparison within a certain time window. Two patches will only be considered
for similarity rating, if they were submitted within a time window of one year.
In the evaluation, we show that this covers 99.5\% of all patches. Secondly,
two patches can not be similar if they do not modify at least one common file.
This fact can be used for further optimisation: we select only pairs of
patches, that modify at least one \emph{similar} file.

In addition to that, we first determine clusters of similar patches for emails
($\mathcal{M} \times \mathcal{M}$). At the beginning of the evaluation, every
email is assigned to its own single-element cluster. We successively merge
clusters in an iterative process by comparing representatives of clusters
against each other. A representative of a cluster is the patch with the
youngest submission date. We choose this patch as representative, as it will
have the closest similarity with further revisions, or with the commit in the
repository, if it was integrated.

After the creation of the clusters for emails, representatives of those
clusters are compared against the commits in the repository.

\subsection{Working with mailing list data}
The first step of the process is the acquisition of mailing list data. This can
be done by subscribing to mailing lists and collecting data; historic data can
be received from archives of a list.

The second step is to filter relevant emails containing patches and to convert
them to a unified format that can be used for further
processing~\cite{bird2006mining}. There are plenty of methods how a user may
send a patch, or how the mail user agent (MUA) may treat the message. Our
parser is able to identify the most commonly used methods. It respects patches
in attachments, (mis-)encoding and different mail parts.

\section{Evaluation}
\label{sec:evaluation}
The results of a heuristic method depend on the chosen set of
parameters. In the following, we identify significant predictors from
the available set of tuneables, and further evaluate the algorithms
accuracy for the optimal choice.

To establish a ground truth, we chose a one-month time window (May 2012, a
typical month of Linux kernel development without any exceptional events) of
the high-volume Linux Kernel Mailing
List\footnote{\url{linux-kernel@vger.kernel.org}} (LKML). We extracted mails
with patches and manually compared them against a three month time window in
the repository in an elaborate and time-consuming task using interactive
support of our tool. The creation of a sound ground truth requires
domain-specific knowledge to judge the relationship of patches, which is
available by some of the authors' active involvement in the respective
communities.

We then analysed the same data with our automated approach, under permutation
of parameters in a reasonable range, as shown in Table~\ref{tab:params}. Prior
to choosing the exact parameter ranges, we performed a coarse-grained analysis
to roughly estimate the influence of parameters. The chosen domains result in
803682 different analysis runs.

In the observed time frame, the list received 16431 emails. Among
these, we identified 5470 containing patches (33.3\%). Assisted by our tool
(and supported by an interactive interface that ensures a swift workflow), the
patches were compared against all commits between Linux kernel versions
\texttt{v3.3} and \texttt{v3.6} (34732 commits).  Those commits are within the
time window 2012-03-18\,--\,2012-09-30 (see Section~\ref{sec:patch_integration}
for a justification of this choice).

The ground truth consists of 3852 clusters of patches, where 2525 clusters are
linked to at least one commit in the repository. 990 clusters contain more than
one email (e.g., multiple revisions of a patch), 394 clusters more than two
emails, and 154 more than three emails. 1712 clusters contain exactly one
email, which means the changes were immediately accepted after their initial
submission without further refinements.

The ground truth is then compared against all clusters from the permutation of
parameters as shown in Table~\ref{tab:params}. In other words, we compare the
ground truth against the 803682 results of our tool.
\begin{table}
\centering
\caption{Set of parameters result used for evaluation}
\begin{tabular}{llll}
	\toprule
	Parameter & Description & Interval & Step \\
	\midrule
	tf  & threshold filename    & $[0.60, 1.00]$ & $0.05$ \\
	th  & threshold heading     & $[0.15, 1.00]$ & $0.05$ \\
	dlr & diff-length ratio     & $[0.00, 1.00]$ & $0.10$ \\
	w   & message-diff weight   & $[0.00, 1.00]$ & $0.10$ \\
	ta  & threshold auto-accept & $[0.60, 1.00]$ & $0.01$ \\
	\bottomrule
\end{tabular}
\label{tab:params}
\end{table}

\subsection{External Evaluation}

External evaluation methods quantify the similarity of two
clusterings~\cite{inforetr}. While there are many standard evaluation methods
available, the correct choice relies on the structure of the
clustering~\cite{amigo2009comparison}. In contrast to typical clustering
problems where a large number of elements (e.g., \emph{documents}) is
distributed to a small number of clusters (e.g., \emph{document types}), our
problem entails a large number of clusters (similar patches) with only few
elements (patch revisions and commits in repositories) per cluster. This
inherently implies a considerable number of ``true negatives'' (TN), since two
randomly chosen elements are assigned to two distinct clusters with high
probability. For a sufficiently large number of clusters, any random clustering
will exhibit a high number of TNs.

\begin{figure}
	\centering
	\resizebox{1.0\linewidth}{!}{
		\input{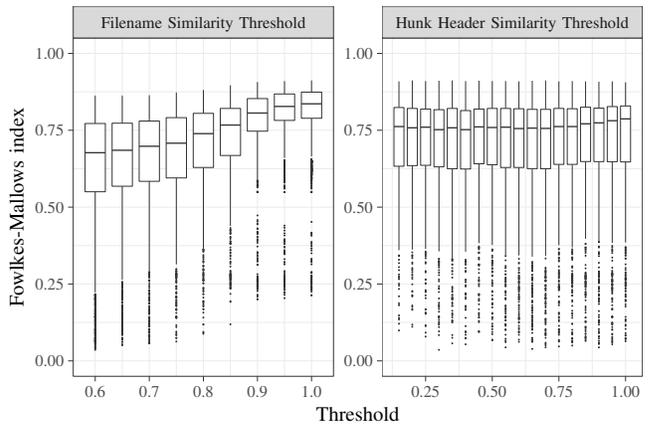}
	}
	\caption{Boxplot of irrelevant parameters: filename and hunk header
		 threshold have no substantial influence.}
	\label{fig:tfth}
\end{figure}

\begin{figure}[h]
	\centering
	\resizebox{0.99\linewidth}{!}{
		\input{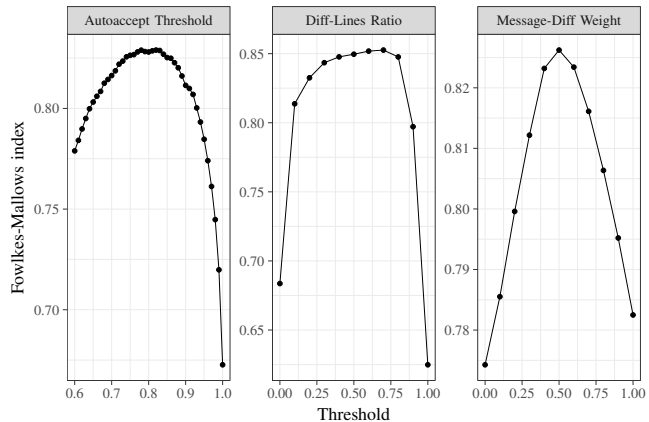}
	}
	\caption{Illustration of the influence of autoaccept
          threshold, diff-length ratio and the message-diff weight
          (connecting lines in all figures are used to
          guide the eye).}
	\label{fig:tadlrw}
\end{figure}

We tested several external evaluation methods for their suitability: mutual
information score~\cite{inforetr}, purity~\cite{inforetr}, V-measure~\cite{vm},
and the Fowlkes-Mallows index~\cite{fowlkes-mallows}. Purity is not suitable
for our problem because it intrinsically produces good results for large
cluster count. A high number of clusters always implies good
purity~\cite{inforetr}.  The V-measure is the harmonic mean of two other
measures, completeness and homogeneity, and also produces good results when
many clusters are present. We consequently choose the Fowlkes-Mallows index,
since it is not sensitive to the number of TN, and shows robust results for
clusterings with a high number of clusters. The Fowlkes-Mallows FM index is
defined as
\begin{equation}
	\text{FM} = \sqrt{
		\frac{\text{TP}}{\text{TP} + \text{FP}} \cdot
		\frac{\text{TP}}{\text{TP} + \text{FN}}
	},
\end{equation}
where TP denotes the number of true positives, and FP and FN provide the number
of false postives and negatives, respectively.

A way to confirm the validity and suitability of an index is to compare it
against an unrelated clustering~\cite{inforetr}. Therefore, we compare the
ground truth against a random clustering, while maintaining the structure of
the clustering, that is, the number of clusters and the number of elements per
cluster. Compared against the ground truth, this reveals a bad Fowlkes-Mallows
index of 0.05. Since the results for our analyses lie within the interval
$[0.231, 0.911]$, this indicates a high validity of the chosen index.

To identify parameters with a relevant influence on the result, we compute the
Fowlkes-Mallows index for each of the 803682 clusterings against the ground
truth. This provides a similarity score for clusterings for each combination of
parameters. To draw conclusions on the significance of a parameter, we
selectively observe the distribution of the Fowlkes-Mallows index for each
parameter. Figure~\ref{fig:tfth} illustrates the Fowlkes-Mallows index for
different values of the filename threshold resp. the hunk header threshold. We
found that different settings for tf and th have little influence on the
results. Instead, best results are achieved for the boundary setting 1 in both
cases (we analyse the reason for the behaviour Section~\ref{sec:discussion}).
For the further analysis, we only regard the subset of our results with
$\text{tf} = 1$ and $\text{th} = 1$ due to their lack of significance. This
requires to consider 2662 clusterings.

Figure~\ref{fig:tadlrw} shows the plot of the mean of the Fowlkes-Mallows index
for autoaccept threshold, diff-length ratio and message-diff weight. Having the
filename and hunk header threshold set to 1, our approach performs best with a
autoaccept threshold of 0.82, a diff-length ratio of 0.4 and a message-diff
weight of 0.3. With this combination, it achieves a Fowlkes-Mallows index of
0.911 on the selected time window.

To confirm the universal validity of those parameters for the whole project, we
cross check the parameters with another mailing list: the linux-commits-tip
mailing list. Every patch that is committed to the Linux tip repository is
automatically sent to the linux-commits-tip mailing list~\cite{jiang2013will}
by the tip-bot. In contrast to standard emails, they contain the commit hash
in the corresponding repository in their header. This allows for simple
cross-validation of the best parameter set. The list can be used to prove the
general functioning of the approach, as the analysis should lead to an exact
match of all patches.

Using a sample of 1047 emails from linux-tip-commits ML compared to the
linux-tip-commits repository, we obtain a Fowlkes-Mallows index of 0.988. Some
minor mismatches are caused by very close, but still dissimilar patches that
are erroneously considered similar, and induced by technical corner cases where
the diff for a patch being sent to the mailing list produces different output
as the diff in the repository (e.g., mode-changes of files or moved files). In
sum, there were 1086 TPs, 18 FPs, and 9 FNs. Note that there are more TPs than
actual emails, because some clusters correctly contain more than one email or
more than one commit; a correct cluster with $n$ elements contains $n\choose 2$
TPs. Once more, these numbers underline the high accuracy of our approach.

\subsection{Example: Duration of patch integration}
\label{sec:patch_integration}

Comparing patches is a computationally intensive task. The number of
comparisons can be reduced if potential comparison candidates are restricted to
patches within a certain time window, as less patches are considered for the
eventual cost-intensive comparison. Our tool already provides a set of
qualitative analyses, such as the integration duration of a patch.

To determine the size of this window, we re-run the analysis on the whole LKML
and the whole repository with the determined optimal set of parameters. We
define the time interval between the date of the latest revision of a patch
(i.e., email submission date) and the date of integration in the repository
(i.e., the commit date) as integration duration.

\begin{figure}
	\resizebox{1.0\linewidth}{!}{
		\input{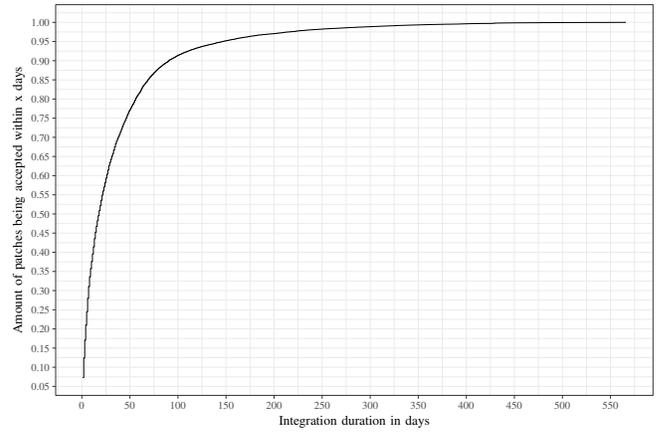}
	}
	\caption{Empirical distribution function of the integration duration of patches on the LKML}
	\label{fig:integration}
\end{figure}

Figure~\ref{fig:integration} shows the empirical distribution function of the
integration duration of all patches of the 99.9\% quantile of all patches.
Interestingly, within the outliers beyond that quantile we found patches that
took indeed five years for integration. 99.5\% of all patches were integrated
within one year, 80\% of all patches within 40 days, 50\% of all patches within
one week.

\subsection{Comparison to other approaches}

In~\cite{german-tracing}, Jiang and colleagues also present a method for
mapping patches on mailing lists to repositories. Their Plus-Minus-based
approach assigns each tuple of changed line and filename to a set of ids, where
the id can either be a message ID or a commit hash. They then search for
patches that contain sufficient identical changes. A threshold between $[0, 1]$
determines the fraction of the number of identical changes that needs to be
exceeded if patches are considered similar.

\begin{figure}
	\centering
	\resizebox{1.0\linewidth}{!}{
		\input{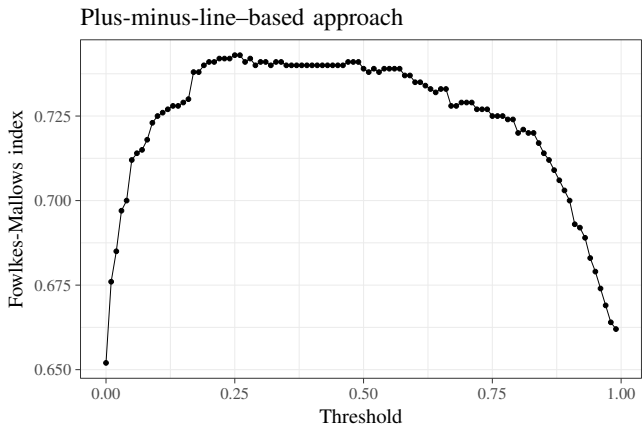}
	}
	\caption{Evaluation of the Plus-Minus-based approach: highest FM index
		 at 0.26, while the threshold only has little influence between
		 $[0.25, 0.74]$}
	\label{fig:pm}
\end{figure}

We used their original implementation to evaluate it against the time window of
our ground truth, and vary their threshold setting in the range $[0, 1]$.
Figure~\ref{fig:pm} shows the results of the analysis. The threshold has no
significant impact on the accuracy within the range $\approx [0.25, 0.75]$. The
best Fowlkes-Mallows index of 0.743 that we could reach with their method is
observed at threshold 0.26.

\section{Discussion}
\label{sec:discussion}

We previously showed the high accuracy of our method, and quantitatively
compared it with other existing techniques methods. We will now turn our
attention to interpreting the meaning of the optimal set of tuneable
parameters, further discuss other methods, and examine the performance (and,
thus, practical applicability) of our approach.

\subsection{Our algorithm}

In Section~\ref{sec:evaluation} we found that both, filename and hunk header
threshold, produce best results for the boundary value 1.00. A filename
threshold of 1.00 implies that patches on the list will not be associated with
a commit in the repository if affected files were renamed between submission
and integration of the patch, and the hunk header threshold of 1.00 disregards
relocations of a hunk within a file. The rationale for these extreme settings
is that both, file moves and relocations within a file, do not occur frequently
in real-world development. It is unlikely that a patch hits this exact window.
While a lower threshold improves recall, it disproportionally decreases
precision since more patches are erroneously considered similar when
relocations occur.

In contrast to filename and hunk header threshold, other parameters
significantly influence the results: auto accept threshold, diff-length ratio
and message diff weight. As expected, too strong or too weak thresholds lead to
over- and underfitting. The diff-length ratio of 0.4 is reasonable because it
allows, for instance, an initial two-line patch to expand into five-line patch
in a future revision, but filters for strongly imbalanced sizes of patches. It
is, for instance, unlikely that a one-line patch will evolve into a 20-line
patch in a future revision. A message-diff weight of 0.3 underlines the
importance to consider both, commit message and diff, with a slight bias
towards the code. It also stresses that involving actual code for analyses is
vital.

\subsection{Plus-Minus-based approach}

While not explicitly mentioned in their paper, the authors
of~\cite{german-tracing} chose a threshold of 0.5 for their algorithm, based on
their experience and intuition\cite{bram-private}. Our evaluation of the
Plus-Minus-based approach shows evidence that this threshold is within a range
where the algorithm performs best.

The authors determine the accuracy of their approach based on the
F-Score, defined as
$F = 2 \cdot \frac{\text{precision}\cdot\text{recall}}{\text{precision}+\text{recall}}$.
It requires knowledge of precision \emph{and} recall. While calculating
precision is straightforward (i.e., counting the number of true and false
positives), a solid ground truth is required to determine the exact recall of
an algorithm, as the recall requires to know the number of false negatives.
They argue that it is hard to determine such a ground truth (a statement that
we fully agree with), and therefore employ the concept of ``relative recall''.
The relative recall incorporates results of the checksum--based technique and
the clone-detection--based technique. The accuracy of these approaches is not
known and therefore relative recall only forms an approximation with unknown
quality. Hence, we think that our determined ground truth leads to more precise
results.

\subsection{Performance}

Performance is an important factor for real world practicability. In
particular, a well-performing implementation is required for the evaluation of
the optimum parameter set, as it requires to run several analyses. Therefore,
we massively parallelise steps of the analysis.

The full analysis of the Linux kernel (v2.6.12\,--\,v4.18 against the whole ML)
with our method requires 13 hours on a machine equipped with two Xeon
E5-2650 processors (20 cores / 40 threads) using the optimal thresholds derived
in Section~\ref{sec:evaluation}. This includes run-once preparation steps like
converting mailing list data to a suitable format, parsing mailing lists for
patches or creating caches.

We were not able to run the full analysis of the Linux kernel with the
plus-minus-line--based approach, because of limitations of their
implementation.

Nonetheless, we found that the plus-minus-line--based approach is considerably
more performant than our approach. For the one-month test set, the approach
takes 80 seconds on the same machine as mentioned before, and only consumes one
single CPU core. Our approach takes between two and eight minutes to analyse
the same set, depending on selected thresholds. The comparison of textual
equivalence used by the plus-minus-line--based technique is less
computation-intensive than our use of Levenshtein string distances.

Yet, our approach is applicable for real world use cases and its best
Fowlkes-Mallows index is 22\% higher than the best score achieved by the
plus-minus-line--based approach.

\section{Threats to Validity}
\subsection{External Validity}
We focus on the Linux kernel for the evaluation, which has strict submission
guidelines, such as requiring detailed commit messages. Patches must be
structured in a fine-grained fashion and must only introduce one small change.
Other projects established different strategies, such as less-verbose commit
messages or larger patches.

Because of this fact, our set of parameters that we found in the evaluation are
therefore thresholds that \emph{suit} Linux, but are not necessarily applicable
to other projects. As a consequence, this demands to repeat the evaluation,
when analysing other projects that the Linux kernel, in order to determine its
proper set of thresholds.

However, numerous other low-level systems that are object of our analyses
adopted the submission guidelines of the Linux kernel that are known as best
practises in the communities. While not mentioned in this paper due to its
length, the same set of parameters lead to high accuracy in other such projects
(e.g., QEMU, Busybox, U-Boot, \dots).

\subsection{Internal Validity}
Other than a perfect gold standard, a manually created ground truth underlies
some uncertainties. The creator may be biased or misjudge decisions, and there
is always a certain degree of subjectivity. The creation of our ground truth
(judging similarity of patches) was carefully done by an experienced developer
with domain-specific knowledge and a track record of active participation in
several open source communities, including the Linux kernel, and we are
confident that our ground truth contains negligible faults.

\subsection{Construct Validity}
Working with mailing lists requires handling noisy data.  Bird et
al.~\cite{bird2006mining} found that 1.3\%~of the Apache HTTP Server Developer
mailing list contains malformed headers.

We need to filter emails on such lists, and consequently use a custom
best-effort parser adapted to handle these difficulties. Since authors may
submit their patches in many ways, finding all patches cannot be guaranteed,
though. Based on the knowledge in the ground truth, the amount of patches that
are not captured is insignificant.  Additionally, the revision control system
git that is widely used for Linux kernel development provides tool support to
prevents common mistakes in email-based patch flows, which reduces the number
of unparseable emails. Following op.~cit., we deem this threat minor.

\section{Related Work II}
Finding similar patches needs to be distinguished from detecting similar code.
\emph{Code clone detection} (CCD) is a well-researched topic mainly driven by
revealing code plagiarism~\cite{cosma2012approach} or redundancy
reduction~\cite{baxter1998clone}. The underlying problems of detecting similar
patches and detecting similar code are related, but differ in one decisive
property: code clone detection analyses a certain \emph{snapshot} of the code,
while detecting similar patches requires analysing a \emph{diff}, which
comprises only fragments detached from the code base. Additionally, a patch
also contains an informal commit message that is not considered by CCD.

Many CCD techniques use language-dependent lexical analysis and analyse
similarities of abstract syntax trees~\cite{deckard, baxter1998clone}.  Since
patches only provide differences between syntactically incomplete fragments of
code, and may also modify non-code artefacts, CCD techniques are typically
inapplicable in our scenario.

Another approach uses locality sensitive hash functions for quantifying code
similarity~\cite{deckard, saebjornsen2009detecting}. Such hash functions
produce similar output for similar input. Arwin et al.\ proposed a
\emph{language independent} approach~\cite{arwin2006plagiarism} that analyses
intermediate code produced by the compiler. This is not applicable to our
problem since the aforementioned analysis of documentation, scripts,
build-system artefacts etc.\ needs to be independent of any language
restrictions.

Bacchelli et al.~\cite{bacchelli2009benchmarking, bacchelli2011miler,
bacchelli2010linking} link emails to source code artefacts in a repository. In
contrast to our work, they focus on discussions and conversations instead of
analysing mails with patches. Naturally, informal conversations have a
different structures than patches.  However, our approach of linking patches on
mailing lists to repositories allows us to transitively link follow-up
discussions of a patch, since the Message-ID of the initial patch remains in
the ``reference header'' of responses.

\section{Conclusion \& Future Work}
\label{sec:conclusion}

The industrial deployment of OSS is often hindered by required certification of
their non-formal development processes according to relevant standards, such as
IEC~61508~\cite{iec61508} for safety-critical industrial, or
ISO~26262~\cite{iso26262} for safety-critical automotive software. Even though
the open and community-driven development process of OSS provides full
traceability of its development, most of the information is not explicitly
contained in the repository, but implicitly hidden in semi-formal discussions
on mailing lists.

We presented a method that is able to reliably link emails with patches to
commits in repositories with high accuracy. Additionally, we formalised the
mathematical background of the problem and identified it as a clustering
problem. Based on this, an elaborate evaluation built upon a solid ground truth
quantifies the high accuracy of our approach. The ground truth and our
framework can be used to evaluate the accuracy of other approaches, and the
fully published framework allows for independent (industrial) evaluation
required in certification efforts.

The evaluation verified that the presented approach performs better than
existing work. For Linux and the LKML, we achieve a 22\% larger Fowlkes-Mallows
index of 0.911 than the best score achieved by the (previously best)
plus-minus-line--based approach.

From the technical and methodological side, future work will focus on improving
the performance of our approach by using hybrid evaluation techniques. This is
intended to combine the performance of fast algorithms with lower accuracy with
the high accuracy of our computationally intensive approach.

Other upcoming work will focus on assessing of non-formal OSS development
processes. Our tool provides the basis for such analyses, as it systematically
makes the history of the process accessible. Its accuracy makes it suitable for
further qualitative software analyses.

\section{Acknowledgements}
We thank Bram Adams for kindly sharing the original implementation of the
plus-minus--based approach~\cite{german-tracing} that made the comparison
against our technique possible. We also thank Julia Lawall and Lukas Bulwahn
for helpful comments on the manuscript.

\clearpage
\printbibliography
\end{document}